\documentclass{eas}
\usepackage{graphicx}
\begin{document}

%%-----------------------------
%%      the top matter
%%-----------------------------
\title{Open clusters in the Gaia-ESO Survey: \\ tracing the chemical history of the Milky Way thin disk} 
\runningtitle{The open clusters in GES}
\author{Laura Magrini}
\author{Sofia Randich}\address{Osservatorio Astrofisico di Arcetri, Largo E. Fermi, 5-I50125-Florence (Italy)}

\begin{abstract}
The Gaia-ESO Survey (GES) is a large public spectroscopic survey that aims at observing with FLAMES@VLT the main stellar components of our Galaxy. 
The study of the population of open clusters is one of the main objectives of GES. 
We present some results from the first 18 months of observations, among them, a preliminary view of the radial metallicity gradient as traced by open cluster data
and a  comparison of  the chemical patterns of clusters located in different parts of the Galactic disk.  
\end{abstract}
\maketitle
%%-----------------------------
%%      your text
%%-----------------------------
\section{Introduction}
The Gaia-ESO Survey (GES) is a large public spectroscopic survey (240+60 nights, from the end of 2011 to the end of 2016) 
that observes with FLAMES@VLT the main stellar components of our Galaxy: the thin and thick disks, the bulge and the halo, and the population of open stellar clusters. More that 10$^5$ stars will be observed at the GES 
completion, mostly at a resolution of R$\sim$20~000 with GIRAFFE and a smaller percentage ($\sim$10$^4$ stars)  with UVES at a higher resolution R=47000. 
The main characteristics and science goals of the Gaia-ESO Survey are described in \cite{gilmore12} and \cite{randich13}.  The main objectives include the study of  Galactic chemo-dynamics, of cluster formation and evolution, and  of stellar evolution. 

\section{The open clusters in GES}
\begin{figure}
   \centering
  \includegraphics[width=0.7\textwidth, angle=270]{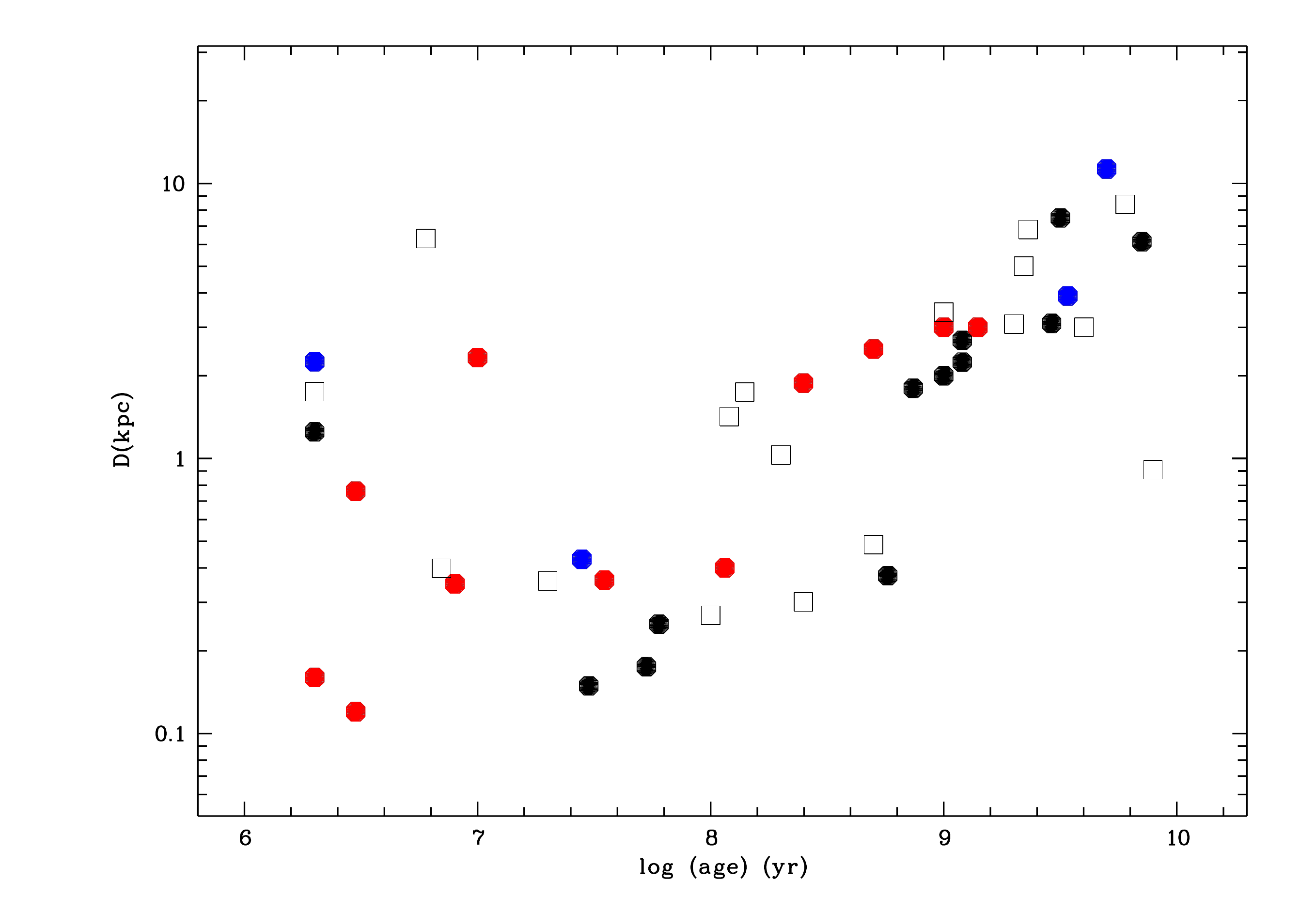}
    \caption{Age (Gyr) versus Distance to the Sun (kpc) of the observed and proposed clusters for the next periods. The filled red circles represent the clusters released in IDR2, in blue in iDR3, and in black in iDR4. The empty squares show the clusters protected for the ESO observational periods P94 and P95. 
}
\label{fig:clusters_ges}
\end{figure}

The GES expects to observe approximatively  70-80 open clusters with the purpose of obtaining a complete 
sampling of the parameter space: ages, distances to the Sun, Galactocentric distances (R$_{\rm GC}$), 
masses, densities, and metallicities. 
The proposed clusters cover large ranges both in ages and distances:
they  extend from very young star forming regions, young clusters with pre-main sequence stars, to intermediate-age and old clusters with more than 0.1~Gyr and 
reaching up to 8~Gyr; the Galactocentric range is well sampled with clusters located in the innermost part of the Galactic disk (R$_{\rm GC} <$~6~kpc) to clusters located in the outskirts of our Galaxy (R$_{\rm GC}\sim$17~kpc).  
In Figure~\ref{fig:clusters_ges} we present the location in the age versus distance (from the Sun) plane of the observed clusters and of those protected  for the next periods.  
The clusters located in the lower-left part of the figure are those mainly involved in the kinematics studies and have important legacy value with GAIA. 
The clusters in the upper part of the plot (D$>$1~kpc)  are instead noteworthy for their application to the study of the radial metallicity gradient in the Galactic thin disk and its evolution with time. 

\section{The radial metallicity gradient as traced by open clusters}
Open clusters are powerful probes of the conditions of the Milky Way Galaxy at all ages and in a wide range of locations in the
disk. Their parameters (age, distance, reddening) 
are promptly available by, e.g.,  isochrone fitting of their colour-magnitude diagrams, and have typical 
uncertainties of the order of 5-10\%. 
They are indeed known to be amongst the most reliable tracers of the Galactic radial metallicity gradient, and, notably, of its time evolution. 
 
The study of the radial metallicity gradient through open clusters sees the first pioneering work at the ends of the 70s,  
when \cite{janes79}, using the photometric metallicity determination of about 40 clusters located within 6~kpc from the Sun, derived, for the first time, 
a radial gradient of [Fe/H], with metallicity decreasing from the inner to the outer parts of the Galaxy at a rate 
d[Fe/H]/dR=$-$0.05~dex. 
During the following decades, many authors have tackled the argument, using gradually more accurate determination of the cluster parameters, 
and, especially, of their metallicity through low, medium and finally high resolution spectroscopy. 
A non-exhaustive list of works dedicated to the study of the Galactic gradient with open clusters includes \cite{friel95}, \cite{friel02}, \cite{yong05}, Sestito et al. (2006, 2008),
\cite{pancino10}, \cite{heiter13}. 
Larger sampled including clusters of different ages allowed also to deal with the subject of the temporal variation of the radial gradient as, e.g., \cite{chen03}, \cite{magrini09}, \cite{jacobson11}, 
\cite{andreuzzi11}, \cite{frinchaboy13}. 
Taking all together the literature results, the main achievements can be summarised in: {\em i)} the {\em global}, i.e. including clusters of all ages, gradient traced by open clusters shows a decreasing metallicity to a {\em breaking radius}, R$_{\rm GC}$$\sim$10-12~kpc, and then  it presents a plateau in metallicity at $\sim$-0.3 to -0.5~dex from the {\em breaking radius} to the most faraway clusters,  i.e. $\sim$20-23~kpc (cf. \cite{pancino10} for a different view);
{\em ii)} there is a notable dispersion in metallicity at any Galactocentric distance, typically of $\sim$0.5~dex; {\em iii)} concerning the time evolution of the gradient, there is a suggestion that younger 
clusters follow a flatter gradient than older clusters up to the breaking radius. However, due to the small number of clusters in different age bins, the slopes of the gradient at different ages are 
still sensitive to the selection of age bins and to the definition of the position of the {\em breaking radius} (cf., e.g.,  Magrini et al. 2009, Jacobson et al. 2011). 

\section{The GES contribution to the Galactic metallicity gradient}
One of the main goals of the GES is the use of open clusters as tracers of the Galactic dick and, in particular,  to define the shape of Galactic metallicity gradient,  to investigate possible 
azimuthal and  vertical gradients, and finally to asses their time evolution.
In this framework, the selection of old and intermediate-age clusters to be observed by GES has been done to pursue also these objectives.
In the first 18 month of observations, we have collected and analysed spectra of several young and old/intermediate-age clusters. 
The recommended stellar parameters and abundances of their stars have been distributed in the first data internal releases. Here we consider the most recent results, i.e. those of {\sc idr2} and {\sc idr3}\footnote{{\sc idr3} is an incremental release, fully consistent with {\sc idr2}} of the UVES spectra. The results used in the present work are those of {\sc wg11}
presented in \cite{smiljanic14}. 
In Table~\ref{tab:clusters} we present the main parameters of the clusters (coordinates, age, R$_{\rm GC}$) and the metallicity derived from GES {\sc idr2} or {\sc idr3}. 
For clusters whose GES data were already published we  use parameters derived in the papers quoted in the last column. For NGC~6705, NGC~4815, and Tr~20 we refer to 
the metallicity of {\sc idr2} as computed, for instance, by \cite{tautvaisiene14}. 
For the remaining clusters, we used literature values for age and R$_{\rm GC}$, and we derived a preliminary membership, based on radial velocities only, 
to compute their mean metallicity [Fe/H] from {\sc idr2} or {\sc idr3}. In the last column we show  the name of the release from which 
the radial velocities and metallicity are taken. 

\begin{figure}
   \centering
  \includegraphics[width=1\textwidth]{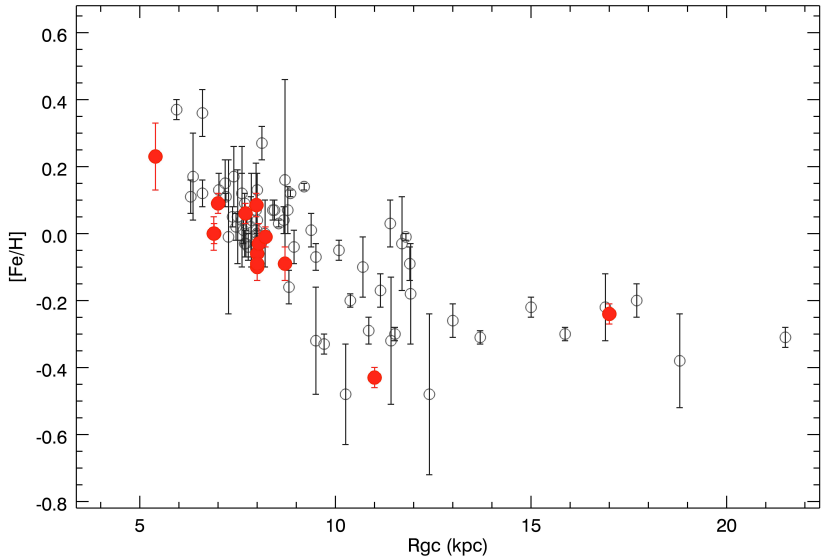}
    \caption{The radial metallicity [Fe/H] gradient as traced by open clusters. Red circles show the open clusters  observed by GES in {\sc idr2} and {\sc idr3}, 
    empty circles are a collection of literature high resolution spectroscopic data from Magrini et al. (2009, 2010).  
}
\label{fig:gradient_ges}
\end{figure}

In Figure~\ref{fig:gradient_ges} we compare the GES results with a collection of high resolution spectroscopic data of open clusters (Magrini et al. 2009, 2010). 
From the Figure we note that, for the literature sample,  the dispersion at each R$_{\rm GC}$ is quite significant, reaching the largest amplitude  $\sim$0.5~dex
around the {\em breaking radius}. Although the selected literature sample consists of only high resolution data, they are possibly affected by differences in the signal-to-noise of different samples, 
in the analysis technique, in the zero point that, altogether, might increase the dispersion. Thus, we might expect that, with a sample of clusters analysed in a fully homogeneous way, the artificial dispersion at each R$_{\rm GC}$
would be reduced and only the dispersion due to real effects would remain. 

The present GES sample of {\sc idr2}/{\sc idr3} is composed by a small number of clusters. However, it is already appreciable that the scatter at any radius is quite reduced. 
The maximum dispersion is obtained at Solar radius ($\sim$0.2~dex) where more clusters were observed. 
However more clusters are necessary to asses the existence of the high dispersion present in the literature data at the 
breaking radius. 
Furthermore, the most distant cluster, Be~25, confirms the existence of a metallicity plateau in the outer Galaxy with important implications for the process of formation of disk galaxies (cf., e.g., 
Bresolin et al. 2009).

\begin{table}
\small
\begin{center}
\caption{Clusters' parameters}
\begin{tabular}{lrllllll}
\hline\hline
{\sc Name} 	&{\sc RA} 	&{\sc dec} 			&{\sc Age} 	&{\sc R$_{\rm GC}$}(a) &[Fe/H] & Ref.	\\
		& \multicolumn{2}{c}{J2000.0} 		        &                       (Gyr) 	&(kpc) & & 	 \\		
\hline\hline
Be~81		&19:01:36	 &	-00:31:00 	    & 	0.86$\pm$0.1 	& 5.4    & +0.23$\pm$0.08 &Magrini et al. (in prep.) \\
NGC 6705 	&18:51:05 &	-06:16:12 	&	0.30$\pm$0.05 & 6.3     & +0.00$\pm$0.05(*)   &Cantat-Gaudin et al.(2014), Tautvaisiene et al.(2014)*\\
Tr~20		&12:39:32 &	-60:37:36 	&	1.50$\pm$0.15 & 6.9    &  +0.10$\pm$0.08(*)  &Donati et al.(2014), Tautvaisiene et al.(2014)*\\
NGC 4815 	&12:57:59 &	-64:57:36 	&	0.57$\pm$0.07 & 6.9    &  -0.01$\pm$0.04(*)  &Friel et al. (2014), Tautvaisiene et al.(2014)*\\
Cha~I		&11:06:48 &	-77:18:00	&	0.002		& 8	    & -0.10$\pm$0.04  &Spina et al.(2014b)\\
$\gamma^2$~Vel&08:09:32&   -47:20:12	&	0.008		& 8	    & -0.06$\pm$0.02  &Spina et al.(2014a)\\
Rho~Oph		&16:28:00 &	-24:25:30	&	0.003		& 7.9	    & -0.09$\pm$0.02		& Spina et al.(in prep.)\\	
NGC~2264	&06:40:58	 &	+09:53:42&	0.003		&8.8	    &-0.09$\pm$0.05		& Spina et al.(in prep.)\\
NGC~2547	&08:10:00	 &	-49:12:00	&	0.003		&8.1	    &-0.03$\pm$0.06		&Spina et al.(in prep.)\\	
NGC~2516	&07:58:04 & 	-60:45:11  & 	0.12	&7.9& +0.08$\pm$0.03& {\sc idr2}\\
IC~4665		&17:46:18	 & 	+15:43:00	&	0.004		&8.2	    &-0.01$\pm$0.03		&{\sc idr3}\\
NGC~2243	&06:29:34	 &	-31:17:00 &	4.7		&11 & -0.43$\pm$0.03	&{\sc idr3}\\			
Be~25		&06:41:00	 &	-16:31:00	&	4.5		&17 &-0.24$\pm$0.03	&{\sc idr3}\\
\hline \hline
\end{tabular}
\label{tab:clusters}
\end{center}
(a) computed with R$_{\odot}$= 8 kpc;
(*) [Fe/H] from Tautvaisiene et al.(2014).
\end{table}

\section{The clusters' chemical patterns}
\begin{figure}
   \centering
  \includegraphics[width=1\textwidth]{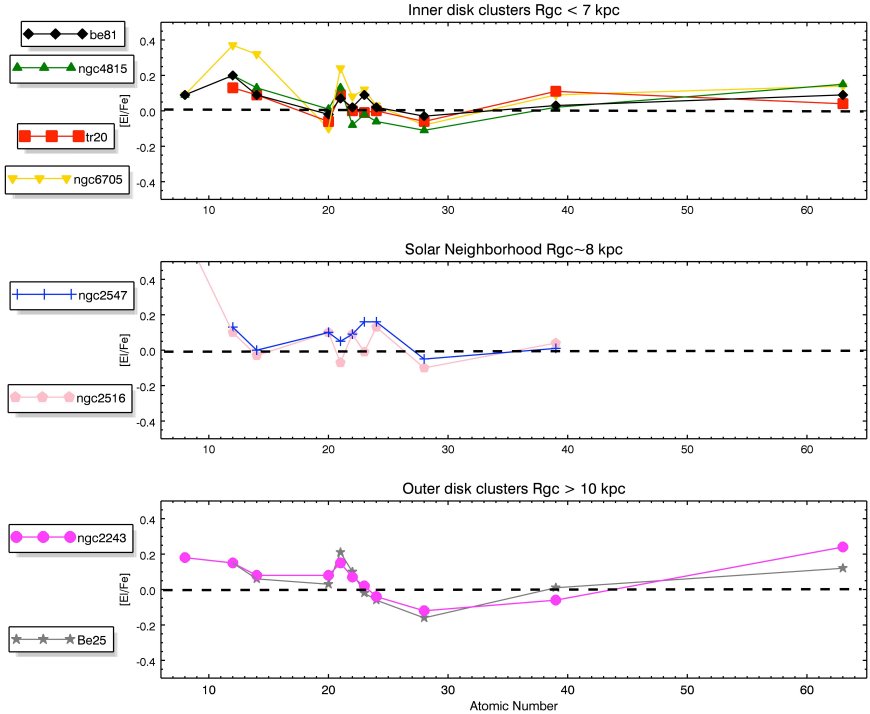}
    \caption{Chemical pattern of old/intermediate-age clusters: atomic numbers (oxygen--8, magnesium--12, silicon--14, calcium--20, scandium--21, titanium--22, vanadium--23, chromium--24, nickel--28,  
yttrium--39, europium--63) vs. average cluster abundances [El/Fe].
    Clusters are divided on the bases of their  R$_{\rm GC}$: in the upper panel the inner disk clusters, in the central panel the solar neighbourhood clusters, and in the 
    lower panel the outer disk clusters. 
}
\label{fig:tagging}
\end{figure}

The high resolution UVES spectra allow to derive the abundances of a large variety of chemical elements, belonging to different nucleosynthesis channels. 
They permit us to compare in a fully consistent way the detailed chemical patterns of open clusters  located in different part of our Galaxy. 
In Figure~\ref{fig:tagging} we show the chemical patterns of mainly old and intermediate-age clusters (only NGC~2547 among the clusters younger than 0.1~Gyr is considered here) of Table~\ref{tab:clusters}. 
In the {\em x} axes we show the atomic numbers  and  in the {\em y} axes the average cluster abundances over iron,  [El/Fe].   
The elements presented in the plot can be grouped in $\alpha$-elements (O, Mg, Si, Ca, Ti) --produced by massive stars in SNII explosions--, iron-peak elements (Sc,Va, Cr, Ni) --produced mainly by SNIa--, and neutron-capture elements (Y, and Eu)--produced by low/intermediate-mass stars (Y) or by massive stars (Eu). 
Although limited by small statistics, we note that groups of clusters located at similar R$_{\rm GC}$ share similar chemical patterns. 
Remarkable differences are seen for some abundance ratios, for instance, in NGC~6705 with respect to the other inner-disk clusters (cf. Magrini et al. 2014). 
However, many similitudes are present in the patterns of the three groups. 
Particularly surprising is the comparison of the two outer disk clusters that, notwithstanding the large difference in R$_{\rm GC}$, show a
striking likeness, a possible signature of the homogeneity, not only in iron, of the outer disk of the Milky Way galaxy. 

\section{Conclusions} 
The open clusters observed during the first 18 months of the Gaia-ESO Survey are showing a great potentiality for the study of the abundance distribution in the Milky Way thin disk. 
They trace a well-defined gradient in the inner part of the disk with a dispersion $<$0.2~dex at each R$_{\rm GC}$. The farthest cluster, Be~25, confirm the existence of an outer 
metallicity plateau. 
The variety of elements available from the UVES analysis allows tracing the chemical patters of clusters located in different regions of the disk. 
Grouping clusters in distance bins has allowed us to discover remarkable similitudes in their chemical patters that are hints of the Galactic radial enrichment history.

%%-----------------------------
%%      your bibliography
%%-----------------------------

\end{document}